\documentclass[12pt]{article} 
\title{A Spacetime for SU(N)}  
\author{{\it Richard Shurtleff~}\thanks{affiliation and mailing 
address: Department of Applied Mathematics and Sciences, 
Wentworth Institute of Technology, 550 Huntington Avenue, 
Boston, MA, USA, ZIP 02115, telephone number: (617) 989-4338, fax 
number: (617) 989-4591 , e-mail address: shurtleffr@wit.edu}} 
\begin{document} 
          
\maketitle 

\begin{abstract} 
The rotations, boosts, and translations in an $N^2$-dimensional spacetime are shown to be related to the fundamental commutators, anticommutators, and Clebsch-Gordan coefficients, respectively, of SU($N$).

\vspace{2cm}


11.30.Cp Lorentz and Poincar\'{e} invariance

 
\vspace{2cm}

\end{abstract}

\pagebreak

\section{Introduction} \label{intro}
Rotations and boosts in $N^2$-dimensional spacetime, $(N^2 - 1)$ dimensional space plus time, are deduced from the commutation and anticommutation relations, respectively, of the generators of the fundamental representations of SU($N$). 

Let $J$ denote a set of matrix generators of SU($N$). The $J$s determine real-valued coefficients $f$ and $d$ in the commutation relations $[J,J]$ = $ifJ$ and the anticommutation relations $\{J,J\}$ = $dJ,$ where indices are suppressed. We show that rotations in SU($N$) spacetime are generated by matrices with components $if$ and boosts are generated by matrices with components $id.$  

Translations are more complicated. One finds that translations connect different representations of SU($N$). When restricted to one irreducible representation (irrep), translations are simply multiplication by unity, trivial. Since $N^2$-dimensional spacetime is an irrep, it follows that vectors  are invariant under translations. But this is not the most interesting possibility.

The translation matrices in a given representation are generated by momentum matrices, just as rotations are generated by angular momentum matrices. Do not confuse momentum matrices with `particle momentum', a vector. These momentum matrices must obey Poincar\'{e} commutation relations with the generators of rotations and boosts. It is shown in this paper that the Poincar\'{e} commutation relations imply that momentum matrices are arrays of Clebsch-Gordan coefficients connecting a rep with the direct sum of two irreps.   

Thus there can be nontrivial translations in SU($N$) spacetime if we consider the reducible rep obtained by combining SU($N$) spacetime with at least one of a select few close `relatives'. These relatives combine spacetime with a fundamental rep. Momentum matrices in SU($N$) spacetime are collections of Clebsch-Gordan coefficients connecting spacetime and its relatives.

The work here generalizes previous work for $N$ = 2 and 3.\cite{N2,N3} The point of view differs from the many other algebraic approaches that generalize four-dimensional spacetime by using SU($N$).\cite{arXiv:math-ph/0703003v2,PhysRevLett56;1532} For one thing, invariants of either quadratic or higher order are ignored. It is known that distance $\sqrt{{x^i}^2},$ $i \in$ $\{1,...,N^2-1\}$ and time $t$ = $x^{N^2}$ are rotation invariants for any $N$ in a suitable realization and it is also known that only four-dimensional spacetime, i.e. SU(2), has ${{x^i}^2} - {t^2}$ invariant under boosts. The invariance of the spacetime interval is a consequence of the simplicity of SU(2) and not a fundamental characteristic to be emulated by higher $N.$ 

The goal here is to show that rotations reflect the algebra of the fundamental commutators, boosts reflect the algebra of the fundamental anticommutators, and translations reflect the algebra of the Clebsch-Gordan coefficients of SU($N$).

\section{A Basis Rep} \label{FunRep}
Consider the set $W$ of $N\times N$ matrices with complex components. Any such matrix $A \in$ $W$ is the sum of a hermitian matrix $(A + A^{\dagger})/2$ and an antihermitian matrix, $(A - A^{\dagger})/2.$ The dagger denotes the hermitian conjugate, the operation combining the transpose and complex conjugation, $A^{\dagger}_{rs}$ = $A_{sr}^{\ast},$ $r,s \in$ $\{1,...,N\}.$

One may choose a set of hermitian matrices $J^{\mu}_{(N)} \in$ $W,$  $\mu \in$ $\{1,...,N^{2}\},$ that spans the set of all hermitian  matrices meaning that any hermitian matrix $B,$ $B^{\dagger}$ = $B,$  can be written as a linear combination of the $J^{\mu}_{(N)}$s with real coefficients. And any anti-hermitian matrix $C,$ $C^{\dagger}$ = $-C,$  can be written as a linear combination of $iJ^{\mu}$s with real coefficients. We have, for hermitian $B$ and anti-hermitian $C,$
\begin{equation} \label{basis}
B = {\beta_{\nu} J^{\nu}_{(N)}} \quad ; \quad  C = {i \gamma_{\nu}  J^{\nu}_{(N)}} \quad ,
\end{equation}
for $N^{2}$ real-valued coefficients $\beta_{\nu}$ and $N^{2}$ real-valued coefficients $\gamma_{\nu}.$ The rule of summing over repeated indices is enforced.

The trace of a matrix $J^{\mu}_{(N)}$ may or may not be zero, but at least one of the matrices $J^{\mu}_{(N)}$ must have a nonzero trace. Suppose exactly one matrix $J^{\mu}_{(N)}$ has a nonzero trace and put that matrix last. One has
\begin{equation} \label{trace}
 {\mathrm{tr}}{(J^{i}_{(N)})} \equiv J^{i}_{(N)\,ss} =  0 \quad ; \quad  {\mathrm{tr}}{(J^{N^2}_{(N)})} = J^{N^2}_{(N)\,ss} \neq  0 \quad ,
\end{equation}
where $i \in$ $\{1,...,N^2-1\}.$  The matrix with nonzero trace, i.e. $J^{N^2}_{(N)},$ need not be the unit matrix.  

A unitary matrix $U \in$ $W,$ $U^{\dagger}U$ = $UU^{\dagger}$ = $\mathbf{1},$ with determinant one, $\det{U}$ = 1, can be written as the matrix exponential of an anti-hermitian matrix ${H}$ with zero trace, i.e. $U$ = $\exp{{H}} $ with ${H}^{\dagger}_{rs}$ = $-{H}^{\ast}_{sr}$ and ${H}_{ss}$ = 0. It follows from (\ref{basis}) that, for some real-valued parameters $\theta_{i},$  $\theta_{i}$ = $\gamma_{i}$ in (\ref{basis}), we have 
$$U = D_{(N)}(\theta)  =  \exp{(i\theta_{i}J^{i}_{(N)})} \quad .$$
There is no $N^2$-component $\theta_{N^2},$ no time component, because $H$ is traceless and $J^{N^2}_{(N)}$ is not traceless. The unitary matrices $U \in$ $W$ with determinant one combined with matrix multiplication form a group, SU($N$), and the matrices $J^{i}_{(N)},$ $i \in$ $\{1,...,N^2-1\},$ form a set of generators of that group. 

The commutators and anticommutators of  hermitian matrices are antihermitian and hermitian, respectively. Thus, by (\ref{basis}), we have two sets of real coefficients, $f^{\mu \nu \sigma}$ and $d^{\mu \nu \sigma},$ 
\begin{equation} \label{Comm1} [J^{\mu}_{(N)},J^{\nu}_{(N)}]  = i f^{\mu \nu \sigma} J^{\sigma}_{(N)}  \quad;
 \end{equation}
\begin{equation} \label{antiComm1} \{J^{\mu}_{(N)},J^{\nu}_{(N)}\}  =  d^{\mu \nu \sigma} J^{\sigma}_{(N)}\quad,
 \end{equation}
 where the commutator is defined by $[J^{\mu},J^{\nu}] \equiv$ $J^{\mu}J^{\nu}-J^{\nu}J^{\mu},$ the anticommutator is defined by $\{J^{\mu},J^{\nu}\} \equiv$ $J^{\mu}J^{\nu}+J^{\nu}J^{\mu},$ and matrix multiplication is understood. 

Associated with the basis set of hermitian matrices $J^{\mu}_{(N)} \in$ $W,$ there is a second basis set of hermitian matrices $\bar{J}^{\mu}_{(\bar{N})} \in$ $W,$ the `anti-rep',
\begin{equation} \label{antiJ}
 \bar{J}^{\mu}_{(\bar{N}) st} \equiv -J^{\mu}_{(N) ts} = -J^{\mu \, \ast}_{(N) st}\quad {\mathrm{(anti\,rep)}} \quad .
\end{equation}
The matrices $\bar{J}^{\mu}_{(\bar{N})}$ are the negative transpose of the $J^{\mu}_{(N)}$ matrices. Since the $J^{\mu}_{(N)}$ are hermitian, the $\bar{J}^{\mu}_{(\bar{N})}$s are also the negative complex conjugate of the $J^{\mu}_{(N)}$ matrices. 

It follows that the commutators and anticommutators of the anti-rep $\bar{J}^{\mu}_{(\bar{N})}$s are 
\begin{equation} \label{Comm1anti} [\bar{J}^{\mu}_{(\bar{N})},\bar{J}^{\nu}_{(\bar{N})}]  = i f^{\mu \nu \sigma} \bar{J}^{\sigma}_{(\bar{N})}  \quad {\mathrm{and}} \quad \{\bar{J}^{\mu}_{(\bar{N})},\bar{J}^{\nu}_{(\bar{N})}\}  = - d^{\mu \nu \sigma} \bar{J}^{\sigma}_{(\bar{N})}\quad,
 \end{equation}
which differ from (\ref{Comm1}) and (\ref{antiComm1}) by the sign of the coefficients $d^{\mu \nu \sigma}.$

 {\it{Remark 2.1.}} Since the commutator has zero trace, $[A,B]_{rr}$ = $A_{rs}B_{sr} - $ $B_{rs}A_{sr} $ = 0, and, by (\ref{trace}), since $J^{N^2}_{(N)}$ is the only matrix $J^{\mu}_{(N)}$ with a non-zero trace, we have $f^{\mu \nu N^2}$ = 0. 
 
 {\it{Remark 2.2.}} The matrix $J^{N^2}$ is the only $J^{\mu}$ with nonzero trace and $J^{N^2}$ need not be the unit matrix~$\mathbf{1}.$ Therefore $J^{N^2}$ need not commute with the $J^{i}$s. Thus $f^{i N^2 k}$ may not be zero. By Remark 2.1 $f^{i k N^2}$ must vanish. It follows that the coefficients $f^{\mu \nu \sigma}$ may not be antisymmetric in $\nu \sigma.$

The generators $J^{i}_{(N)}$ are here called `angular momentum matrices' as well as generators and the $\theta$ parameters are `angles'. The matrix $D_{(N)}(\theta)$ realizes a `rotation' through angle $\theta$ with angular momentum generators $J^{i}_{(N)}.$ The rotation generated by the $\bar{J}^{i}_{(\bar{N})}$s is written $D_{(\bar{N})}(\theta).$

 \section{Poincar\'{e} Commutation Relations} \label{PCR}
 
We define boost matrices $K$ and vector matrices $V.$ Then we determine commutation relations among the $J$s, $K$s, and $V$s that have the characteristics that the commutators $[V^{i},K^{j}]$ are symmetric in $ij$, the $[V^{\mu},K^{j}]$ are sums of $V$s, and the $[V^{\mu},J^{j}]$ are sums of $V$s. These characteristics are shared by 4-D spacetime's Poincar\'{e} commutation relations.

We require the Poincar\'{e} commutation relation $[V^{i},K^{j}]$ be symmetric in $ij.$ The following definitions of $V^{i}$ and $K^{j}$ make the commutator $[V^{i},K^{j}]$ depend on the anticommutators $\{J^{i},J^{j}\},$ which are necessarily symmetric in $ij.$

Define $2N\times2N$ matrices $J_{(2N)}$, $K_{(2N)}$, and $V_{(2N)}$ in block notation as follows
 \begin{equation} \label{anti1} J^{\mu}_{(2N)} = \pmatrix{J^{\mu}_{(N)} && 0 \cr 0 && J^{\mu}_{(N)}}  \quad ; \quad K^{i}_{(2N)} = \pmatrix{+iJ^{i}_{(N)} && 0 \cr 0 && -iJ^{i}_{(N)}} \quad,
 \end{equation}
 \begin{equation} \label{anti2} V^{\mu}_{(2N)} = \pmatrix{0 && V^{\mu}_{+} \cr V^{\mu}_{-} && 0}  = \pmatrix{0 && c_{+}J^{\mu}_{(N)} \cr c_{-}J^{\mu}_{(N)} && 0} \quad ,
 \end{equation}
where $c_{+}$ and $c_{-}$ are complex constants. It follows that
 \begin{equation} \label{Comm4} [V^{\mu}_{(2N)},K^{j}_{(2N)}] =  \pmatrix{0&&-ic_{+}\{J^{\mu}_{(N)},J^{j}_{(N)}\} \cr +ic_{-}\{J^{\mu}_{(N)},J^{j}_{(N)}\}  &&0} = \quad
 \end{equation}
\vspace{0.3cm}
$$ = i d^{\mu j\sigma}\pmatrix{0&&-c_{+}J^{\sigma}_{(N)} \cr +c_{-}J^{\sigma}_{(N)}  &&0} \quad . $$  
One sees that $[V^{i},K^{j}]$ is symmetric in $ij$ by the definition of the coefficients $d^{\mu \nu \sigma}$ in (\ref{antiComm1}).

We also require the commutator $[V^{\mu},K^{j}]$ to be a sum of $V^{\nu}$s. Comparing (\ref{Comm4}) with the definition of the $V^{\mu}_{(2N)}$s in (\ref{anti2}) and noting that some $d^{\mu \nu \lambda}$ are nonzero, one sees that the sign difference in the off-diagonal blocks requires that 
$c_{+}$ = 0 or $c_{-}$ = 0. We have
$$V^{\mu}_{+} = 0 \quad {\mathrm{or}} \quad V^{\mu}_{-} = 0 \quad , $$
so one of the off-diagonal blocks of the $V^{\mu}_{(2N)}$ vanishes.

{\it{Remark 3.1.}} Due to the simplicity of the $d^{\mu \nu \sigma}$ in SU(2), one can arrange for both off-diagonal blocks of $V^{\mu}$ to be nonzero. For example, as often displayed, the Dirac gamma matrices are vector matrices with both off-diagonal blocks nonzero.\cite{DIRAC}

It should be clear that two matrices commute when their nonzero components are relegated to the same off-diagonal block. Since vector matrices have nonzero components confined to the 12-block with $c_{-}$ = 0 and nonzero components confined to the 21-block for $c_{+}$ = 0, it follows that vector matrices commute. Commuting vector matrices qualify as `momentum' matrices and are denoted as `$P^{\mu}.$' Updating the notation, we replace $V$ by $P.$ Thus the cases $c_{-}$ = 0 and $c_{+}$ = 0 produce momentum matrices $P^{(\epsilon_{P})\,\mu}_{(2N)}$ with $\epsilon_{P}$ = $\pm 1,$
\begin{equation} \label{P+P-} V \rightarrow P \quad ; \quad P^{(+)\,\mu}_{(2N)} \equiv \pmatrix{0 && c_{+}J^{\mu}_{(N)} \cr 0 && 0}  \quad ; \quad P^{(-)\,\mu}_{(2N)} \equiv \pmatrix{0 && 0 \cr c_{-}J^{\mu}_{(N)}  && 0} \quad .
 \end{equation}
Clearly the matrices commute,
\begin{equation} \label{PP}[P^{(+)\,\mu}_{(2N)},P^{(+)\,\nu}_{(2N)}] = 0 \quad {\mathrm{and}} \quad [P^{(-)\,\mu}_{(2N)},P^{(-)\,\nu}_{(2N)}] = 0 \quad.
 \end{equation}

Collecting the commutation relations of the matrices $J_{(2N)}$, $K_{(2N)}$, and $P^{(\epsilon_{P})}_{(2N)}$ we have 
$$ [J^{i},J^{j}]  = i f^{ijk} J^{k}  \quad ; \quad [J^{i},K^{j}]  = i f^{ijk} K^{k}  \quad ; \quad [K^{i},K^{j}]  =  -i f^{ijk} J^{k} \quad
$$
\begin{equation} \label{Pcomm1} [P^{(\epsilon_{P})\,\mu},J^{j}]  =  i f^{\mu jk} P^{(\epsilon_{P})\,k}  \; ; \; [P^{(\epsilon_{P})\,\mu},K^{j}] = -\epsilon_{P} i d^{\mu j\sigma} P^{(\epsilon_{P})\,\sigma}  \; ; \;   [P^{(\epsilon_{P})\,\mu},P^{(\epsilon_{P})\,\nu}] = 0 
  \quad ,
\end{equation}
for $\mu, \nu, \sigma \in$ $\{1,...,N^2\},$ $i,j,k \in$ $\{1,...,N^2 - 1\},$ and $\epsilon_{P}$ = $\pm 1.$ Some of these are (\ref{Comm4}) and (\ref{PP}), while the rest follow from (\ref{Comm1}) and (\ref{antiComm1}). They make up the Poincar\'{e} commutation relations for SU($N$).

\section{SU($N$)-spacetime.} \label{SUNspacetime}

We obtain the generators of rotations and boosts, the $N^2$-dimensional matrices $J_{(N^2)}$ and $K_{(N^2)},$ that satisfy the Lorentz subset of the Poincar\'{e} commutation relations for SU($N$). 
 
Consider the Poincar\'{e} commutation relations that are in the form $[P,A]$ = $aP.$ One can show that the coefficients $a$ obey the same commutation relations as the matrices $A.$\cite{N2,N3} 

Thus the matrices $J^{j}_{(N^2)}$ and $K^{j}_{(N^2)}$ are just the coefficients of the $P^{(\epsilon_{P})\,\sigma}$s on the right sides of the $[P^{(\epsilon_{P})\,\mu},J^{j}]$ and $[P^{(\epsilon_{P})\,\mu},K^{j}]$ commutation relations (\ref{Pcomm1}), 
\begin{equation} \label{JKN2} 
{J^{i}_{(N^2)\, {\mu \nu}}}  =  if^{\mu i \nu} \quad ; \quad {K^{(\epsilon_{P}) \, i}_{(N^2)\, \mu\nu}}  =  -\epsilon_{P} i d^{\mu i \nu} \quad,
\end{equation}
where the sum $i f^{\mu jk} P^{(\epsilon_{P})\,k}$ in (\ref{Pcomm1}) equals $i f^{\mu j\nu} P^{(\epsilon_{P})\,\nu}$ because $ f^{\mu j N^2}$ = 0.
Since $[P^{(\epsilon_{P})\,\mu},P^{(\epsilon_{P})\,\nu}]$ = $0,$ the coefficients on the right all vanish and the matrices $P^{\nu}_{(N^2)}$ vanish. Momentum matrices and translations are determined another way in the following section.

One can show by (\ref{Pcomm1}) that the matrices $J^{i}_{(N^2)}$ and $K^{(\epsilon_{P}) \, j}_{(N^2)}$ satisfy the following subset of the SU($N$)-Poincar\'{e} commutation relations,
\begin{equation} \label{Lorentz}
 [J^{i},J^{j}]  = i f^{ijk} J^{k}  \quad ; \quad [J^{i},K^{(\epsilon_{P}) \, j}]  = i f^{ijk} K^{(\epsilon_{P}) \, k}  \quad ; \quad [K^{(\epsilon_{P}) \, i},K^{(\epsilon_{P}) \, j}]  =  -i f^{ijk} J^{k} \quad ,
\end{equation}
which are the {\it{Lorentz}} commutation relations for SU($N$).

As an immediate consequence of (\ref{JKN2}) and (\ref{Lorentz}) we have
\begin{equation} \label{JJJJ} [J^{i},J^{j}]  = J^{j}_{(N^{2})\, ik} J^{k}  \quad ; \quad [J^{i},K^{(\epsilon_{P}) \, j}]  = J^{j}_{(N^{2})\, ik} K^{(\epsilon_{P}) \, k}  \quad ; \quad [K^{(\epsilon_{P}) \, i},K^{(\epsilon_{P}) \, j}]   =  -J^{j}_{(N^{2})\, ik} J^{k} \quad .  
\end{equation}
And the $[P,J]$ and $[P,K]$ Poincar\'{e} commutation relations (\ref{Pcomm1}) can be rewritten in a similar fashion. This will be helpful in the next section.

Let 
\begin{equation} \label{Dtrans}
D^{\epsilon_{P}}_{(N^2)}(\theta,\phi) = \exp{(i\phi_{i}K^{(\epsilon_{P}) \, j}_{(N^2)})}\exp{(i\theta_{i}J^{i}_{(N^2)})}
\end{equation}
 be the $N^2$-dimensional Lorentz transformation matrix for a rotation through angle $\theta$ = $\{\theta_{1},...,\theta_{N^2-1}\}$ followed by a boost through $\phi$ = $\{\phi_{1},...\phi_{N^2-1}\}.$ The SU($N$)-Lorentz transformation $D^{\pm}_{(N^2)}$ acts on $N^2$-dimensional vectors such as coordinates $\{x^{i},x^{N^2}\},$ ordered sets of real numbers.

\section{Translation Generators - Momentum Matrices} \label{MomMat}
In this section, we show that momentum matrices are arrays of Clebsch-Gordon coefficients relating certain reps of SU($N$).

Suppose we combine two irreducible representations of SU($N$), labeled $A$ and $B,$ with generators $J^{i}_{(A)\, a a_1}$ and $J^{i}_{(B)\, b b_1}$ as follows
\begin{equation} \label{JAB}
	J^{i}_{(A,B)jk} = J^{i}_{(A,B)a_{1}b_{1},a_{2}b_{2}} = J^{i}_{(A)a_{1}a_{2}}\delta^{b_{1}}_{b_{2}} + \delta^{a_{1}}_{a_{2}}J^{i}_{(B)b_{1}b_{2}} \quad
\end{equation}
and also
\begin{equation} \label{KAB}
	K^{i}_{(A,B)jk} = K^{i}_{(A,B)a_{1}b_{1},a_{2}b_{2}} = -i \left(J^{i}_{(A)a_{1}a_{2}}\delta^{b_{1}}_{b_{2}} - \delta^{a_{1}}_{a_{2}}J^{i}_{(B)b_{1}b_{2}}\right) \quad,
\end{equation}
where $\delta^{a_{1}}_{a_{2}}$ is unity for $a_1$ = $a_2$ and zero otherwise. 

In these expressions the single indices $j$ and $k$ on the left are identified with a prescribed ordering of the double indices $(a_{1},b_{1})$ and $(a_2,b_2),$ respectively, on the right. Thus the matrices ${J^{i}_{(A,B)}}$ and ${K^{i}_{(A,B)}}$ are square matrices with a dimension equal to the product of the dimensions of reps $A$ and $B.$ 

Based on the fact that $J^{i}_{(A)}$ and $J^{i}_{(B)}$ satisfy the fundamental $[J,J]$ commutation relations (\ref{Comm1}), one can show that $J^{i}_{(A,B)} $ and $ K^{i}_{(A,B)}$ obey the Lorentz relations (\ref{Lorentz}). 

As a special case consider (\ref{JKN2}), the angular momentum and boost matrices $J^{i}_{(N^2)}$ and $K^{(\epsilon_{P}) \, i}_{(N^2)},$ $\epsilon_{P}$ = $\pm 1,$ i.e. the rep for SU($N$) spacetime. One can show that $J^{i}_{(N^2)}$ and $iK^{(\epsilon_{P}) \,i}_{(N^2)}$ are similar to the fundamental rep ${J}^{i}_{(N)}$ and the fundamental anti-rep ${\bar{J}}^{i}_{(\bar{N})}$ in the combinations $J^{i}_{(N)} \pm {\bar{J}}^{i}_{(\bar{N})},$
\begin{equation} \label{JN2}
	S^{\mu}_{\sigma}J^{i}_{(N^2)\sigma \nu} = \left(J^{i}_{(N)\mu_{1} \rho_{1}}\delta^{\mu_{0}}_{\rho_{0}} + \delta^{\mu_{1}}_{\rho_{1}}
{ {\bar{J}^{i}} }_{(\bar{N})\mu_{0} \rho_{0}} \right) S^{\rho}_{\nu} \quad
\end{equation}
and also
\begin{equation} \label{KN2}
S^{\mu}_{\sigma}K^{(\epsilon_{P}) \,i}_{(N^2)\sigma \nu} = -\epsilon_{P} i \left(J^{i}_{(N)\mu_{1} \rho_{1}}\delta^{\mu_{0}}_{ \rho_{0}} - \delta^{\mu_{1}}_{\rho_{1}}{ {\bar{J}^{i}} }_{(\bar{N})\mu_{0}\rho_{0}} \right) S^{\rho}_{\nu} \quad,
\end{equation}
where $S$ is a similarity matrix. The single index $\mu$ appears on the left and double indices $\mu_{1}\mu_{0}$ appear on the right. A convenient correspondence is given by
\begin{equation} \label{mumu1mu0}
\mu = N(\mu_{1}-1)+\mu_{0} \quad ; \quad \rho = N(\rho_{1}-1)+\rho_{0} \quad,
\end{equation}
with $\mu \in$ $\{1,...,N^2\}$ and $\mu_{1},\mu_{0} \in$ $\{1,...,N\}.$ Likewise for $\rho.$ A proof of (\ref{JN2}) and (\ref{KN2}) can be found in the Appendix.

By (\ref{JN2}) and (\ref{KN2}), the SU($N$) spacetime rep has $(A,B)$ = $(N,\bar{N})$ for $\epsilon_{P}$ = $+1$ and  $(A,B)$ = $(\bar{N},N)$ for $\epsilon_{P}$ = $-1.$ We turn back now to the general case.

To obtain momentum matrices, we work in the $(A,B)\oplus (C,D)$ rep and define the following matrices 
 \begin{equation} \label{ABCD1} J^{i}_{(ABCD)} \equiv \pmatrix{J^{i}_{(A,B)} && 0 \cr 0 && J^{i}_{(C,D)}}  \quad ; \quad K^{i}_{(ABCD)} \equiv \pmatrix{K^{i}_{(A,B)} && 0 \cr 0 && K^{i}_{(C,D)}} \quad,
 \end{equation}
 \begin{equation} \label{ABCD2} P^{\mu}_{(ABCD)} = \pmatrix{0 && P_{+}^{\mu} \cr P_{-}^{\mu} && 0}  \quad .
 \end{equation}
Compare (\ref{ABCD1}) and (\ref{ABCD2}) with (\ref{anti1}) and (\ref{anti2}). Both here and there, the $J$s and $K$s are block diagonal while the $P$s have off-diagonal blocks. 

We want to find matrices $P^{\mu}_{(ABCD)}$ that, with the known matrices $J^{i}_{(ABCD)}$ and $K^{i}_{(ABCD)},$ obey the Poincar\'{e} commutation relations (\ref{Pcomm1}). For brevity, we discuss only $\epsilon_{P}$ = $+1.$ By (\ref{JKN2}) we have $[P,J]$ = $ifP$ = $J_{(N^2)}P$ and $[P,K]$ = $idP$ = $K_{(N^2)}P.$  And, by (\ref{JN2}) and (\ref{KN2}), $J_{(N^2)}$ and $K_{(N^2)}$ are related to $J^{i}_{(N)} \pm \bar{J}^{i}_{(\bar{N})}$ by a similarity transformation $S.$ Putting all this together, one finds equations that $P^{\mu}_{(ABCD)}$ must obey. 

The $P^{\mu}_{(ABCD)}$s are in block form (\ref{ABCD2}) with nonzero blocks $P_{+}$ and  $P_{-}.$ After a straightforward calculation, one finds $P_{+}$ block equations,
\begin{equation} \label{ABCD3} \tilde{P}^{\lambda_{1}\lambda_{0}}_{+\,absd}J^{i}_{(C)\,sc}  = (J^{i}_{(N)\,\lambda_{1} \rho_{1}}\delta^{a}_{r} + \delta^{\lambda_{1}}_{\rho_{1}} J^{i}_{(A)\,ar})\tilde{P}^{\rho_{1}\lambda_{0}}_{+ \,rbcd}
  \quad , \quad [\epsilon_{P} = +1] \quad
 \end{equation}
$$ \tilde{P}^{\lambda_{1}\lambda_{0}}_{+\,abcs}J^{i}_{(D)\,sd}  = (\bar{J}^{i}_{(\bar{N})\,\lambda_{0} \rho_{0}}\delta^{b}_{r} + \delta^{\lambda_{0}}_{\rho_{0}} J^{i}_{(B)\,br})\tilde{P}^{\lambda_{1}\rho_{0}}_{+ \,arcd}
 \quad ,  \quad [\epsilon_{P} = +1] \quad $$ where single and double indices are related by (\ref{mumu1mu0}) and we define $\tilde{P}$ by
\begin{equation} \label{SimP} \tilde{P}^{\lambda_{1}\lambda_{0}} = \tilde{P}^{\lambda} = S^{\lambda}_{\mu}P^{\mu} \quad,
\end{equation}
where  $S^{\lambda}_{\mu}$ is the similarity transformation in (\ref{JN2}) and (\ref{KN2}). Thus the $C$-rep is related to the $A$-rep and the fundamental rep ($N$), while $D$ is related to $B$ and the anti-rep ($\bar{N}$). 

In terms of rotations $D(\theta)$ = $\exp{(i\theta_{i}J^{i})},$ (\ref{ABCD3}) gives
\begin{equation} \label{ABCD4} \tilde{P}^{\lambda_{1}\lambda_{0}}_{+\,absd} D^{(C)}_{sc}(\theta) = D^{(N\otimes A)}_{\lambda_{1}a, \rho_{1}r}(\theta) \tilde{P}^{\rho_{1}\lambda_{0}}_{+\,rbcd}  \quad \quad [\epsilon_{P} = +1] \quad
\end{equation}
$$  \tilde{P}^{\lambda_{1}\lambda_{0}}_{+\,abcs} D^{(D)}_{sd}(\theta) = D^{(\bar{N}\otimes B)}_{\lambda_{0}b, \rho_{0}r} (\theta)\tilde{P}^{\lambda_{1}\rho_{0}}_{+\,arcd}   \quad \quad [\epsilon_{P} = +1] \quad  $$
Clearly, the components of $\tilde{P}_{+}$ are the Clebsch-Gordan coefficients relating the $C$-rep of SU($N$) with the $N \otimes A$ direct product rep and the $D$-rep with the $\bar{N} \otimes B$ rep.\cite{Hamermesh} Therefore the components of $\tilde{P}_{+}$ are products of Clebsch-Gordan coefficients,
\begin{equation} \label{ABCD5} \tilde{P}^{\lambda}_{+\,abcd}  = \tilde{P}^{\lambda_{1}\lambda_{0}}_{+\,abcd} = k_{+} (NA\lambda_{1}a \mid Cc)(\bar{N}B\lambda_{0}b \mid Dd) \quad ,  \quad [\epsilon_{P} = +1] \quad
 \end{equation}
where $k_{+}$ is a constant and we use (\ref{mumu1mu0}) to relate the $\lambda$s, i.e. $\lambda$ = $N(\lambda_{1}-1)+\lambda_{0}.$  For the $\tilde{P}_{-}$ block, we find that 
\begin{equation} \label{ABCD6} \tilde{P}^{\lambda}_{-\,cdab}  = \tilde{P}^{\lambda_{1}\lambda_{0}}_{-\,cdab} = k_{-} (NC\lambda_{1}c \mid Aa)(\bar{N}D\lambda_{0}d \mid Bb) \quad .  \quad [\epsilon_{P} = +1] \quad
 \end{equation}
Thus momentum matrices $\tilde{P}^{\mu}$ are arrays of Clebsch-Gordan coefficients. 

For $\epsilon_{P}$ = $-1$ exchange fundamental reps $N$ and $\bar{N}$ in the above discussion.

In order for the momentum matrices to commute, $[\tilde{P}^{\mu}_{\pm},\tilde{P}^{\nu}_{\pm}]$ = 0, we make one of the off-diagonal blocks vanish, i.e. $k_{+} $ = 0 or $k_{-} $ = 0,
 \begin{equation} \label{ABCD6a}   {P}^{\mu}_{+\,abcd}  = 0  \quad {\mathrm{or}} \quad {P}^{\mu}_{-\,cdab}  = 0 \quad .
 \end{equation}
By the off-diagonal structure of definition (\ref{ABCD2}), translations generated with $\tilde{P}^{\mu}_{+}$ change quantities in $(A,B)$-space and leaves quantities transforming with $(C,D)$ invariant. And $\tilde{P}^{\mu}_{-}$ changes $(C,D)$ quantities and leaves $(A,B)$ quantities invariant.

 {\it{Remark 5.1}} One can build bigger matrices by adding irreducible reps, $(A,B)\oplus(C,D)\oplus(E,F)...$. Then the momentum matrices have nonzero blocks whenever $A,C,E...$ are related by a direct product with the fundamental rep $(N)$ as in (\ref{ABCD4}) or when $B,D,F,...$ are related by a direct product with the fundamental anti-rep $(\bar{N})$. 

When applied to SU($N$) spacetime, we have $(A,B)$ = $(N,\bar{N})$ for $\epsilon_{P}$ = $+1,$ as discussed above in this section. If we combine SU($N$) spacetime with another rep $(C,D)$ we get $(A,B)\oplus (C,D)$ = $(N,\bar{N})\oplus (C,D).$ As just shown, we get momentum matrices with the $P_{+}$ block nontrivial when $C \in$ $N \otimes A$ = $N \otimes N$ and/or $D \in$ $\bar{N} \otimes B$ = $\bar{N} \otimes \bar{N}.$ And we get momentum matrices with the $P_{-}$ block nontrivial when $A$ = $N \in$ $N \otimes C$ and/or $B$ = $\bar{N} \in$ $\bar{N} \otimes D.$ By `$ H \in G$' we mean the rep $H$ is included in the sum of irreducible reps that make up the rep $G.$

{\it{Remark 5.2}} Consider the translations generated by the momentum matrices with a nonzero $P^{\mu}_{+}$ block. These translations change SU($N$) spacetime vectors in a way that depends on quantities in the appended $(C,D)$ space. Thus if a particle's momentum is a vector in SU($N$) spacetime, then the particle's momentum changes upon translation in a way dependent on quantities in the appended $(C,D)$ space, i.e. force. And translations generated by the $P_{-}$ block change vectors in $(C,D)$ space. The translations generated by the $P_{-}$ block leave SU($N$) spacetime vectors unchanged.

Translations have been shown in this section to be generated by momentum matrices in the form of arrays of Clebsch-Gordan coefficients, thereby completing the task proposed at the start of this paper.

\appendix

\section{Proof of a Theorem} \label{Proof}

In this appendix, we prove the theorem that there exists a similarity transformation $S$ satisfying (\ref{JN2}), $SJ_{(N^2)}$ = $(J_{(N)} + \bar{J}_{(\bar{N})})S,$ and (\ref{KN2}), $SK_{(N^2)}$ = $-i(J_{(N)} - \bar{J}_{(\bar{N})})S.$ First we prove the theorem for special bases and then extend the result to general bases by considering the transformation from one basis to another.

For some bases the similarity transformation $S^{\lambda}_{\sigma}$ is just a reorganization of the basis matrices $J^{\sigma}_{(N)\, mn}.$ The similarity transformation $S^{\lambda}_{\sigma}$ has two indices each running from 1 to $N^2,$ $\lambda,\sigma \in$ $\{1,...,N^2\},$ while $J^{\sigma}_{mn}$ has indices $m$ and $n$ each running from 1 to $N.$ If we make one index $\lambda$ out of the two indices, e.g. $\lambda$ = $N(m-1)+n,$ then we can have $S^{\lambda}_{\sigma}$ =  $J^{\sigma}_{(N)\,mn}$ = $J^{\sigma}_{(N)\, \lambda}.$ 

Start with the fundamental commutation and anticommutation relations (\ref{Comm1}) and (\ref{antiComm1}),
$$ [J^{\mu}_{(N)},J^{\nu}_{(N)}]_{mn}  = i f^{\mu \nu \sigma} J^{\sigma}_{(N) \, mn}  \quad ; \quad
 \{J^{\mu}_{(N)},J^{\nu}_{(N)}\}_{mn}  =  d^{\mu \nu \sigma} J^{\sigma}_{(N)\, mn} \quad . \quad {\mathrm{[Eqns. (3);(4)]}} 
$$
The left sides may be expanded using the fact that the anti-rep matrices are the negative transpose of the $J_{(N)}$s, i.e. ${\bar{J}}^{\sigma}_{(\bar{N})\, mn}$ = $-J^{\sigma}_{(N)\, nm}.$ One finds that the left sides are
$$ [J^{\mu}_{(N)},J^{\nu}_{(N)}]_{mn}  =  J^{\mu}_{(N)\,ms}J^{\nu}_{(N)\,sn} - J^{\nu}_{(N)\,mt}J^{\mu}_{(N)\,tn} = 
\left(J^{\mu}_{(N)\,ms}\delta^{t}_{n} + \delta^{s}_{m}{\bar{J}}^{\mu}_{(\bar{N})\,nt}\right) J^{\nu}_{(N)\,st} \quad ; \quad 
$$
 \begin{equation} \label{34left} \{J^{\mu}_{(N)},J^{\nu}_{(N)}\}_{mn}   =  J^{\mu}_{(N)\,ms}J^{\nu}_{(N)\,sn} + J^{\nu}_{(N)\,mt}J^{\mu}_{(N)\,tn} = 
\left(J^{\mu}_{(N)\,ms}\delta^{t}_{n} - \delta^{s}_{m}{\bar{J}}^{\mu}_{(\bar{N})\,nt}\right) J^{\nu}_{(N)\,st} \quad.
\end{equation}

On the right sides of  (\ref{Comm1}) and (\ref{antiComm1}), we make the assumptions that the coefficients $f^{\mu \nu \sigma}$ are antisymmetric in $\nu\sigma,$ $f^{\mu \nu \sigma}$ = $-f^{\mu \sigma \nu}$ and the $d^{\mu \nu \sigma}$s are symmetric, $d^{\mu \nu \sigma}$ = $d^{\mu \sigma \nu}.$ A set of matrices $J^{\sigma}_{mn}$ satisfying these assumptions is displayed below. Then in such a special basis by the definitions of $J_{(N^2)}$ and $K_{(N^2)}$ in (\ref{JKN2}), we have for the right sides
$$ i f^{\mu \nu \sigma} J^{\sigma}_{(N) \, mn} = i f^{\sigma \mu \nu} J^{\sigma}_{(N) \, mn} = J^{\sigma}_{(N) \, mn} i f^{\sigma \mu \nu} =  J^{\sigma}_{(N) \, mn} J^{\mu}_{(N^2)\sigma \nu}	\quad ;		$$
\begin{equation} \label{34right} d^{\mu \nu \sigma} J^{\sigma}_{(N) \, mn} = d^{\sigma \mu \nu} J^{\sigma}_{(N) \, mn} = J^{\sigma}_{(N) \, mn} d^{\sigma \mu \nu} = \epsilon_{P} i J^{\sigma}_{(N) \, mn} K^{\epsilon_{P}\mu}_{(N^2)\sigma \nu}	\quad .		
\end{equation}
By including a time component for $K^{\epsilon_{P}\mu}_{(N^2)},$ i.e. $\mu$ = $N^2,$ we generalize (\ref{JKN2}) and prove a more general theorem than is needed.

Using the new expressions for the left and right sides of (\ref{Comm1}) and (\ref{antiComm1}), we get
$$ J^{\sigma}_{(N) \, mn} J^{\mu}_{(N^2)\sigma \nu} = \left(J^{\mu}_{(N)\,ms}\delta^{t}_{n} + \delta^{s}_{m}{\bar{J}}^{\mu}_{(\bar{N})\,nt}\right) J^{\nu}_{(N)\,st} \quad ;	 $$
\begin{equation} \label{34leftright} \epsilon_{P} i J^{\sigma}_{(N) \, mn} K^{\epsilon_{P}\mu}_{(N^2)\sigma \nu} = \left(J^{\mu}_{(N)\,ms}\delta^{t}_{n} - \delta^{s}_{m}{\bar{J}}^{\mu}_{(\bar{N})\,nt}\right) J^{\nu}_{(N)\,st} \quad . 
\end{equation}
Now, with 
\begin{equation} \label{SJ} m, \, n \rightarrow \lambda_{1}, \, \lambda_{0}  \quad ; \quad S^{\lambda}_{\sigma} =   J^{\sigma}_{(N)\,mn} \quad ,
\end{equation}
 where $\lambda$ = $N(m-1)+n$ = $N(\lambda_{1}-1)+\lambda_{0},$ we have
$$ S^{\lambda}_{\sigma} J^{\mu}_{(N^2)\sigma \nu} = \left(J^{\mu}_{(N)\,\lambda_{1} \rho_{1}}\delta^{\rho_{0}}_{\lambda_{0}} + \delta^{\rho_{1}}_{\lambda_{1}}{\bar{J}}^{\mu}_{(\bar{N})\,\lambda_{0}\rho_{0}}\right) S^{\rho}_{\nu} \quad ;	 $$
\begin{equation} \label{SimJK}  S^{\lambda}_{\sigma} K^{\epsilon_{P}\mu}_{(N^2)\sigma \nu} = -\epsilon_{P} i \left(J^{\mu}_{(N)\,\lambda_{1} \rho_{1}}\delta^{\rho_{0}}_{\lambda_{0}} - \delta^{\rho_{1}}_{\lambda_{1}}{\bar{J}}^{\mu}_{(\bar{N})\,\lambda_{0}\rho_{0}}\right) S^{\rho}_{\nu} \quad , 
\end{equation}
where $\rho$ = $N(s-1)+t$ = $N(\rho_{1}-1)+\rho_{0}.$ Thus the theorem is shown for the special reps with completely antisymmetric $f^{\mu \nu \sigma}$s and completely symmetric $d^{\mu \nu \sigma}.$

One such basis is well known.\cite{specialREP} Let $T^{ab}$ be the $N\times N$ matrix with zero for all components except the $ab$th component which is one. We write $T^{ab}_{mn}$ = $\delta^{ab}_{mn},$ where $\delta$ is one when the upper sequence of indices $ab...$ is equal to the lower sequence of indices $mn...$ and zero otherwise. Define the basis matrices $J^{\epsilon ab}_{(N)}$ by
  $$   J^{+ ab}_{(N)}  = \frac{1}{2}\left( T^{ab} + T^{ba} \right)  \quad  {\mathrm{;}} \quad J^{- ab}_{(N)}  = \frac{-i}{2}\left( T^{ab} - T^{ba} \right) \quad 1 \leq a<b \leq N $$ 
\begin{equation} \label{AJN}\quad J^{\,0 ab}_{(N)}  = \frac{\delta^{a}_{b}}{\sqrt{2(a^2-a)}}\left( \sum^{a-1}_{n=1}T^{nn} -(a-1) T^{aa} \right)  \quad a,b \in \{2,...,N\} \quad {\mathrm{;}} \quad J^{\,0 11}_{(N)}  = \frac{1}{\sqrt{2N}}{\mathbf{1}} \quad ,
 \end{equation}
where the boldface ${\mathbf{1}}$ indicates the $N\times N$ identity matrix, ${\mathbf{1}}_{mn}$ = $\delta^{m}_{n}.$ 

There are $N(N-1)/2$ + $N(N-1)/2$ + $N-1$ + $1$ = $N^2$ matrices $J^{\epsilon ab}_{(N)}.$ The $J^{\epsilon ab}_{(N)}$ are hermitian and all but the last, $J^{011}_{(N)},$ are traceless. Let $\sigma$ be the position of $J^{\epsilon ab}_{(N)}$ in the above list (\ref{AJN}). We write 
$$J^{\sigma}_{(N)} \in \{J^{+ ab}_{(N)},J^{- ab}_{(N)},J^{0 ab}_{(N)},J^{011}_{(N)}\} \quad ,$$ 
where $\sigma \in$ $\{1,...,N^2\}.$ One can show that the $J^{\sigma}_{(N)}$s obey the following trace identity,
\begin{equation} \label{trace1} {\mathrm{tr}}(J^{\mu}_{(N)}J^{\nu}_{(N)}) = \frac{\delta^{\mu}_{\nu}}{2} \quad ,
\end{equation}
which, with the fundamental commutators and anticommutators (\ref{Comm1}) and (\ref{antiComm1}), gives
\begin{equation} \label{fdTRACE} f^{\mu \nu \lambda} = -2i{\mathrm{tr}}([J^{\mu}_{(N)},J^{\nu}_{(N)}]J^{\lambda}_{(N)}) \quad {\mathrm{and}} \quad d^{\mu \nu \lambda} = 2{\mathrm{tr}}(\{J^{\mu}_{(N)},J^{\nu}_{(N)}\}J^{\lambda}_{(N)}) \quad .
\end{equation}
Clearly, $f^{\mu \nu \lambda}$ is antisymmetric in $\mu \nu$ and $d^{\mu \nu \lambda}$ is symmetric in $\mu \nu$ because the commutator and anticommutator are antisymmetric and symmetric respectively.

Since the trace of a matrix product $AB$ is independent of the order, ${\mathrm{tr}}(AB)$ = $A_{ms}B_{sm}$ = $B_{sm}A_{ms}$ =  ${\mathrm{tr}}(BA),$ one finds that
$$ f^{\mu \nu \lambda} = -2i{\mathrm{tr}}([J^{\mu}_{(N)},J^{\nu}_{(N)}]J^{\lambda}_{(N)}) = -2i{\mathrm{tr}}(J^{\mu}_{(N)}J^{\nu}_{(N)}J^{\lambda}_{(N)}-J^{\nu}_{(N)}J^{\mu}_{(N)}J^{\lambda}_{(N)}) = $$ $$ -2i{\mathrm{tr}}(J^{\lambda}_{(N)}J^{\mu}_{(N)}J^{\nu}_{(N)}-J^{\mu}_{(N)}J^{\lambda}_{(N)}J^{\nu}_{(N)}) = +2i{\mathrm{tr}}([J^{\mu}_{(N)},J^{\lambda}_{(N)}]J^{\nu}_{(N)}) = -f^{\mu \lambda \nu} \quad ,$$
which shows that $f^{\mu \nu \lambda}$ is antisymmetric in $\nu \lambda.$ Since $f^{\mu \nu \lambda}$ is antisymmetric in $\mu \nu$ as shown already and $f^{\mu \nu \lambda}$ is now shown to be antisymmetric in $\nu \lambda,$ $f^{\mu \nu \lambda}$ is completely antisymmetric. By the same reasoning, one can show that $d^{\mu \nu \lambda}$ is completely symmetric in this special rep. 

By (\ref{SimJK}), since the $f^{\mu \nu \lambda}$ are completely antisymmetric and the $d^{\mu \nu \lambda}$s are completely symmetric, the theorem holds for this special representation.

The traceless matrices in (\ref{AJN}), $J^{i}_{(N)}$  with $i \in$ $\{1,...,N^2-1\},$  form a basis of all hermitian $N\times N$ traceless matrices. When we include $J^{N^2}_{(N)}$ = $J^{011}_{(N)},$ we have a basis $J^{\mu}_{(N)}$ with $\mu \in$ $\{1,...,N^2\}$ for all hermitian $N\times N$ matrices traceless or not.

Any other basis $J^{\prime \, i}_{(N)},$ $i \in$ $\{1,...,N^2-1\},$ of traceless hermitian $N\times N$ matrices can be extended to a set of basis matrices $J^{\prime \, \mu}_{(N)}$ of  hermitian $N\times N$ matrices, traceless or not, by including a hermitian matrix  $J^{\prime \, N^2}_{(N)}$ with nonzero trace. We want just one matrix with nonzero trace so that there are $N^2-1$ traceless hermitian matrices to generate the matrices of SU($N$), see Section 2.

It follows that there exists an $N^2 \times N^2$ matrix $R$ with real components such that
\begin{equation} \label{RJ}
J^{\prime \, \mu}_{(N)} = R^{\mu}_{\sigma}J^{\sigma}_{(N)}\quad,
\end{equation}
with indices $\mu,\sigma$ taking values in $\{1,...,N^2\}.$ By (\ref{trace1}), we have $R^{\mu}_{\lambda}$ = $2{\mathrm{tr}}(J^{\prime \, \mu}_{(N)}J^{\lambda}_{(N)}).$ The $J^{\prime \, i}_{(N)},$ $i \in$ $\{1,...N^2-1\},$ are all traceless, so $R^{i}_{N^2}$ = 0 and $J^{\prime \, i}_{(N)} = R^{i}_{j}J^{j}_{(N)}.$ The primed basis is a basis for the unprimed $J^{\mu}_{(N)}$s, so $R$ has an inverse, $R^{-1}.$ 

Applied to the anti-rep, the fact that $R$ is real implies that 
\begin{equation} \label{JbarPRIME}\bar{J}^{\prime \, \mu}_{(\bar{N})\, st} = -{J}^{\prime \, \mu \, \ast}_{(N) st} = -\left(R^{\mu}_{\sigma}{J}^{\sigma}_{(\bar{N})\,st}\right)^{\ast} = R^{\mu}_{\sigma}\bar{J}^{\sigma}_{(\bar{N})\,st} \quad .
\end{equation}
Also, since $R^{i}_{N^2}$ = 0, we have $\bar{J}^{\prime \, i}_{(\bar{N})\, st}$ =  $R^{i}_{j}\bar{J}^{j}_{(\bar{N})\,st},$ with $i,j \in$ $\{1,...,N^2-1\}.$

Consider (\ref{Comm1}) and (\ref{antiComm1}), i.e. the commutators $[J^{\mu}_{(N)},J^{\nu}_{(N)}]$ = $if^{\mu \nu \sigma}J^{\sigma}_{(N)}$ and anticommutators $\{J^{\mu}_{(N)},J^{\nu}_{(N)}\}$ = $d^{\mu \nu \sigma}J^{\sigma}_{(N)}.$ By (\ref{RJ}), the commutators and anticommutators of the primed basis have coefficients 
\begin{equation} \label{fdprime}
f^{\prime \, \mu \nu \lambda} = R^{\mu}_{\sigma}R^{\nu}_{\rho}f^{\sigma \rho \tau}R^{-1 \, \tau}_{\lambda} \quad {\mathrm{and}} \quad d^{\prime \, \mu \nu \lambda} = R^{\mu}_{\sigma}R^{\nu}_{\rho}d^{\sigma \rho \tau}R^{-1 \, \tau}_{\lambda} \quad.
\end{equation}
Since $if$ and $id$ are $J_{(N^2)}$ and $K_{(N^2)}$ within a sign, it follows that 
\begin{equation} \label{fdprime1}
{J^{\prime \, \nu}_{(N^2)\, \mu \lambda}} = R^{\mu}_{\sigma}R^{\nu}_{\rho}{J^{ \rho}_{(N^2)\, \sigma \tau}}R^{-1 \, \tau}_{\lambda} \quad {\mathrm{and}} \quad {K^{\epsilon_{P} \, \prime \, \nu}_{(N^2)\, \mu \lambda}} = R^{\mu}_{\sigma}R^{\nu}_{\rho}{K^{\epsilon_{P} \, \rho}_{(N^2)\, \sigma \tau}}R^{-1 \, \tau}_{\lambda} \quad.
\end{equation}
Finally, we define $S^{\prime \, \mu}_{\nu}$ 
\begin{equation} \label{fdprime2}
S^{\prime \, \mu}_{\nu} = S^{\mu}_{\sigma} R^{-1 \, \sigma}_{\nu} \quad
\end{equation}
so that, by (\ref{SimJK}), (\ref{RJ}) and (\ref{JbarPRIME}), we have
\begin{equation} \label{JN2appPRIME}
	S^{\prime \, \lambda}_{\sigma}{J^{\prime \, \mu}_{(N^2)\, \sigma \nu}} = \left(J^{\prime \, \mu}_{(N)\lambda_{1} \phi_{1}}\delta^{\lambda_{0}}_{\phi_{0}} + \delta^{\lambda_{1}}_{\phi_{1}}
{ {\bar{J}^{\prime \, \mu}} }_{(\bar{N})\lambda_{0} \phi_{0}} \right) S^{\prime \, \phi}_{\nu} \quad
\end{equation}
and also
\begin{equation} \label{KN2appPRIME}
S^{\prime \, \lambda}_{\sigma}K^{(\epsilon_{P}) \, \prime \, \mu}_{(N^2)\, \sigma \nu} = -\epsilon_{P} i \left(J^{\prime \, \mu}_{(N)\lambda_{1} \phi_{1}}\delta^{\lambda_{0}}_{\phi_{0}} - \delta^{\lambda_{1}}_{\phi_{1}}
{ {\bar{J}^{\prime \, \mu}} }_{(\bar{N})\lambda_{0} \phi_{0}} \right) S^{\prime \, \phi}_{\nu}   \quad,
\end{equation}
where $\lambda$ = $N(\lambda_{1}-1)+\lambda_{0}$ and $\phi$ = $N(\phi_{1}-1)+\phi_{0}.$ The inclusion of time components $\mu$ = $N^2$ makes the theorem here more general than (\ref{JN2}) and (\ref{KN2}).

Thus the theorem is satisfied. For any set of basis matrices $J^{\prime \, \mu}_{(N)}$ that has just one matrix, i.e. $J^{\prime \, N^2}_{(N)},$ with a nonzero trace, there is a similarity transformation $S^{\prime \, \lambda}_{\sigma}$ satisfying (\ref{JN2appPRIME}) and (\ref{KN2appPRIME}).

\pagebreak

\section{Exercises and Problems} \label{Pb}

The on-line article has a problem set.

\vspace{0.3cm}
\noindent 1. Show that the matrices ${J^{i}_{(A,B)}}$ and ${K^{i}_{(A,B)}},$ (\ref{JAB}) and (\ref{KAB}), satisfy the Lorentz commutation relations. [Hint: Use the commutation relations for the $A$ and $B$ reps, $[J^{\mu}_{(A)},J^{\nu}_{(A)}]$ = $i f^{\mu \nu \sigma} J^{\sigma}_{(A)}$ and $[J^{\mu}_{(B)},J^{\nu}_{(B)}]$ = $i f^{\mu \nu \sigma} J^{\sigma}_{(B)},$ or some other way.]

\vspace{0.3cm}
\noindent 2. Show that (\ref{ABCD3}) follows from (\ref{JKN2}) and (\ref{JN2}) - (\ref{ABCD2}).

\vspace{0.3cm}
\noindent 3.  (a) Show that the square of the distance in SU($N$)-spacetime ${x^{i}}^{2}$ is invariant under rotations.   (b) Show that time, i.e. $x^{N^2},$ is invariant under rotations. Do we need a special basis with completely antisymmetric $f^{\mu \nu \lambda}$ and completely symmetric $d^{\mu \nu \lambda}?$ [Hint: Use the antisymmetry of $f^{i jk}$ in $ij$ originating in the commutator $[J^{i}_{(N)},J^{j}_{(N)}].$ And the $N^2$ components of $J^{i}_{(N^2)}$ vanish because the $f^{ijN^2}$ are zero.]

\vspace{0.3cm}
\noindent 4. (a) Consider the symmetric triple product. Operate on the first two factors and then on the second two factors. By (\ref{Comm1}), we have $J^{(\lambda}_{(N)}J^{\mu}_{(N)}J^{\nu)}_{(N)}$ =  $J^{(\mu}_{(N)}J^{\lambda}_{(N)}J^{\nu)}_{(N)}$ + $i f^{(\lambda \mu \sigma} J^{\sigma}_{(N)}J^{\nu)}_{(N)}$ = $J^{(\lambda}_{(N)}J^{\nu}_{(N)}J^{\mu)}_{(N)}$ + $i f^{(\mu \nu \sigma} J^{\lambda)}_{(N)}J^{\sigma}_{(N)},$ where parentheses indicate the sum over cyclic permutations, e.g. $(123)$ = $123+231+312.$ (b) Now use (\ref{antiComm1}). (c) Rearranging indices $(\mu \nu \lambda)$ =  $(\lambda \mu \nu)$ in (a) and (b), we get
\begin{equation} \label{idff} f^{\lambda \mu \sigma}f^{\sigma \nu \tau} + f^{\mu \nu \sigma}f^{\sigma \lambda \tau} + f^{\nu \lambda \sigma}f^{\sigma \mu \tau}   =  0 \quad
 \end{equation}
\begin{equation} \label{iddf} d^{\lambda \mu \sigma}f^{\sigma \nu \tau} + d^{\mu \nu \sigma}f^{\sigma \lambda \tau} + d^{\nu \lambda \sigma}f^{\sigma \mu \tau}   =  0 \quad.
 \end{equation}
Compare these identities to (\ref{JKN2}) and (\ref{Lorentz}).

\vspace{0.3cm}
\noindent 5. (a) Find $f^{\mu \nu \lambda}$ and $d^{\mu \nu \lambda}$ for SU(2) using the rep in Appendix A. Get the generators $J_{(N^2)}$ = $J_{(4)}$ and $K_{(N^2)}$ = $K_{(4)}$ for SU(2) spacetime, i.e. 4-d spacetime. Find the rotation matrix for $\theta_{i}$ with just one nonzero component, say $\theta_{1},$ and find the boost matrix when $\phi_{i}$ has just $\phi_{1}$ nonzero. (b)~Show that the $f$s and $d$s in (a) are completely antisymmetric and symmetric, respectively. (c) Find $S^{\lambda}_{\sigma}$ =  $J^{\sigma}_{(2)\,mn}$ = $J^{\sigma}_{(2)\, \lambda}.$ (d) Repeat for SU(3).  

\vspace{0.3cm}
\noindent 6. For the basis-to-basis transformation matrix $R$ in the Appendix, show that  we have
$ R^{N^2}_{N^2}$ = ${\mathrm{tr}}(J^{\prime \, N^2}_{(N)})/{\mathrm{tr}}(J^{N^2}_{(N)}) \, ,$
where ${\mathrm{tr}}(M)$ indicates the trace of the matrix $M,$ ${\mathrm{tr}}(M) \equiv$ $M_{ss}.$

\vspace{0.3cm}
\noindent 7. Does the expression $\det{(S)}$ = $i^{N(N-1)/2}/2^{N^2/2}$ give the determinant of $S^{\lambda}_{\sigma}$ =   $J^{\sigma}_{(N)\,mn},$
where $\lambda$ = $N(m-1)+n,$  for all $N >$ 1 in the special representation (\ref{AJN}) of SU($N$) in Appendix~A? 

\end{document}